\pdfoutput=1

\documentclass{ws-procs961x669}

\makeatletter
\let\jnl@style=\rm
\def\ref@jnl#1{{\jnl@style#1}}

\def\nar{\ref@jnl{NewAR}}
\def\apj{\ref@jnl{ApJ}}                 
\def\apjl{\ref@jnl{ApJ}}                
\def\apjs{\ref@jnl{ApJS}}               
\def\aap{\ref@jnl{A\&A}}                
\def\nat{\ref@jnl{Nature}}              
\def\mnras{\ref@jnl{MNRAS}}             
\def\pasp{\ref@jnl{PASP}}               
\def\physrep{\ref@jnl{Phys.~Rep.}}
\def\araa{\ref@jnl{ARA\&A}}             
\def\ssr{\ref@jnl{Space~Sci.~Rev.}}         
\def\aj{\ref@jnl{AJ}}       
\def\apss{\ref@jnl{Ap\&SS}}             
\def\jqsrt{\ref@jnl{J. Quant. Spectrosc. Radiative Transfer}}             
\def\prl{\ref@jnl{Phys. Rev. Lett.}} 
\def\actaa{\ref@jnl{Acta Astronomica}}  
\def\prd{\ref@jnl{Physical Review D}}  
\def\memras{\ref@jnl{Mem. RAS}} 
\def\na{\ref@jnl{New Astronomy}}

\makeatother

\begin{document}

\title{The role of the magnetic fields in GRB outflows}

\author{Jordana-Mitjans, N.$^1$\footnote{E-mail: N.Jordana@bath.ac.uk}; Mundell, C. G.$^1$; Kobayashi, S.$^2$; Smith, R. J.$^2$; Guidorzi, C.$^{3,4,5}$; Steele, I. A.$^2$; Shrestha, M.$^2$; Gomboc, A.$^6$; Marongiu, M.$^{7}$; Martone, R.$^{3}$; Lipunov, V.$^8$; Gorbovskoy, E. S.$^8$; Buckley, D. A. H.$^9$; Rebolo, R.$^{10}$; Budnev, N. M.$^{11}$}

\address{ $^1$Department of Physics, University of Bath, Claverton Down, Bath, BA2 7AY, UK\\
$^2$Astrophysics Research Institute, Liverpool John Moores University, 146 Brownlow Hill, Liverpool, L3 5RF, UK\\
$^{3}$Department of Physics and Earth Science, University of Ferrara, via Saragat 1, I-44122, Ferrara, Italy\\
$^{4}$INFN -- Sezione di Ferrara, Via Saragat 1, 44122 Ferrara, Italy\\
$^{5}$INAF -- Osservatorio di Astrofisica e Scienza dello Spazio di Bologna, Via Piero Gobetti 101, 40129 Bologna, Italy\\
$^{6}$Center for Astrophysics and Cosmology, University of Nova Gorica, Vipavska 13, 5000 Nova Gorica, Slovenia\\
$^{7}$INAF -- Osservatorio Astronomico di Cagliari - via della Scienza 5 - I-09047 Selargius, Italy\\
$^8$Lomonosov Moscow State University, SAI, Physics Department, 13 Univeristetskij pr-t, Moscow 119991, Russia\\
$^9$South African Astronomical Observatory PO Box 9, Observatory 7935, Cape Town, South Africa\\
$^{10}$Instituto de Astrof\'isica de Canarias (IAC), Calle V\'ia L\'actea s/n, E-38200 La Laguna, Tenerife, Spain\\
$^{11}$Applied Physics Institute, Irkutsk State University, 20, Gagarin blvd, 664003, Irkutsk, Russia}

\begin{abstract}

Gamma-ray bursts (GRBs) are bright extragalactic flashes of gamma-ray radiation and briefly the most energetic explosions in the Universe. Their catastrophic origin ---the merger of compact objects or the collapse of massive stars--- drives the formation of a newborn compact remnant (black hole or magnetar) that powers two highly relativistic jets. As these jets continue to travel outwards, they collide with the external material surrounding the dying star, producing a long-lasting afterglow that can be seen across the entire electromagnetic spectrum, from the most energetic gamma-ray emission to radio wavelengths. But how can such material be accelerated and focused into narrow beams? The internal shock model proposes that repeated collisions between material blasted out during the explosion can produce the gamma-ray flash. The competing magnetic model credits primordial large-scale ordered magnetic fields that collimate and accelerate the relativistic outflows. To distinguish between these models and ultimately determine the power source for these energetic explosions, our team studies the polarization of the light during the first minutes after the explosion (using novel instruments on fully autonomous telescopes around the globe) to directly probe the magnetic field properties in these extragalactic jets. This technology allowed the detection of highly polarized optical light in GRB 120308A\cite{2013Natur.504..119M} and confirmed the presence of mildly magnetized jets with large-scale primordial magnetic fields in a reduced sample of GRBs (e.g. GRB 090102 \cite{2009Natur.462..767S}, GRB 110205A\citep{2017ApJ...843..143S}, GRB 101112A\citep{2017ApJ...843..143S}, GRB 160625B\cite{2017Natur.547..425T}). Here we discuss the observations of the most energetic and first GRB detected at very high TeV energies, GRB 190114C\cite{2020ApJ...892...97J}, which opens a new frontier in GRB magnetic field studies suggesting that some jets can be launched highly magnetized and that the collapse and destruction of these magnetic fields at very early times may have powered the explosion itself. Additionally, our most recent polarimetric observations of the jet of GRB 141220A\cite{2021MNRAS.505.2662J} indicate that, when the jetted ejected material is decelerated by the surrounding environment, the magnetic field amplification mechanisms at the front shock ---needed to generate the observed synchrotron emission--- produce small magnetic domains. These measurements validate theoretical expectations and contrast with previous observations that suggest large magnetic domains in collisionless shocks (i.e. GRB 091208B\cite{2012ApJ...752L...6U}). 

\end{abstract}

\bodymatter

\vspace{20px}

Gamma-ray bursts (GRBs) are briefly the most intense sources of gamma-ray photons in the Universe to the extent that they outshine any other kind of gamma-ray emitter present in the sky map. The amount of energy released in a matter of milliseconds to hundreds of seconds provides a unique opportunity to study physics in extreme environments (e.g., GRB central engines, jet composition, energy dissipation, acceleration, collimation, shock physics) and test the physical models regarding the nature of their progenitors\cite{2011CRPhy..12..206Z, 2015PhR...561....1K}. Additionally, their brightness and cosmological distances make them useful for probing the composition of the early Universe\cite{2009Natur.461.1258S,2011ApJ...736....7C}.

GRBs with long duration (typically $> 2\, $s; see Ref.~\citenum{1993ApJ...413L.101K}) are related to the death of massive stars as single collapsars \citep{1993ApJ...405..273W,1999ApJ...524..262M} or in binary systems \citep{2007PASP..119.1211F}, and short GRBs ($<2\,$s) to mergers from an old population of degenerate compact stars. Possible merger candidates include neutron star binaries \citep{1986ApJ...308L..43P,1989Natur.340..126E} or binaries containing a stellar black hole and a neutron star companion \citep{1991AcA....41..257P,1992ApJ...395L..83N,2006ARA&A..44..507W}. Due to their inspiral and coalescence, they are emitters of gravitational waves \citep{2017PhRvL.119p1101A} and they are also promising candidate sources of high-energy neutrinos \citep{2017ApJ...848L...4K}.

\begin{figure}[ht!]
\centering
\includegraphics[width=80mm]{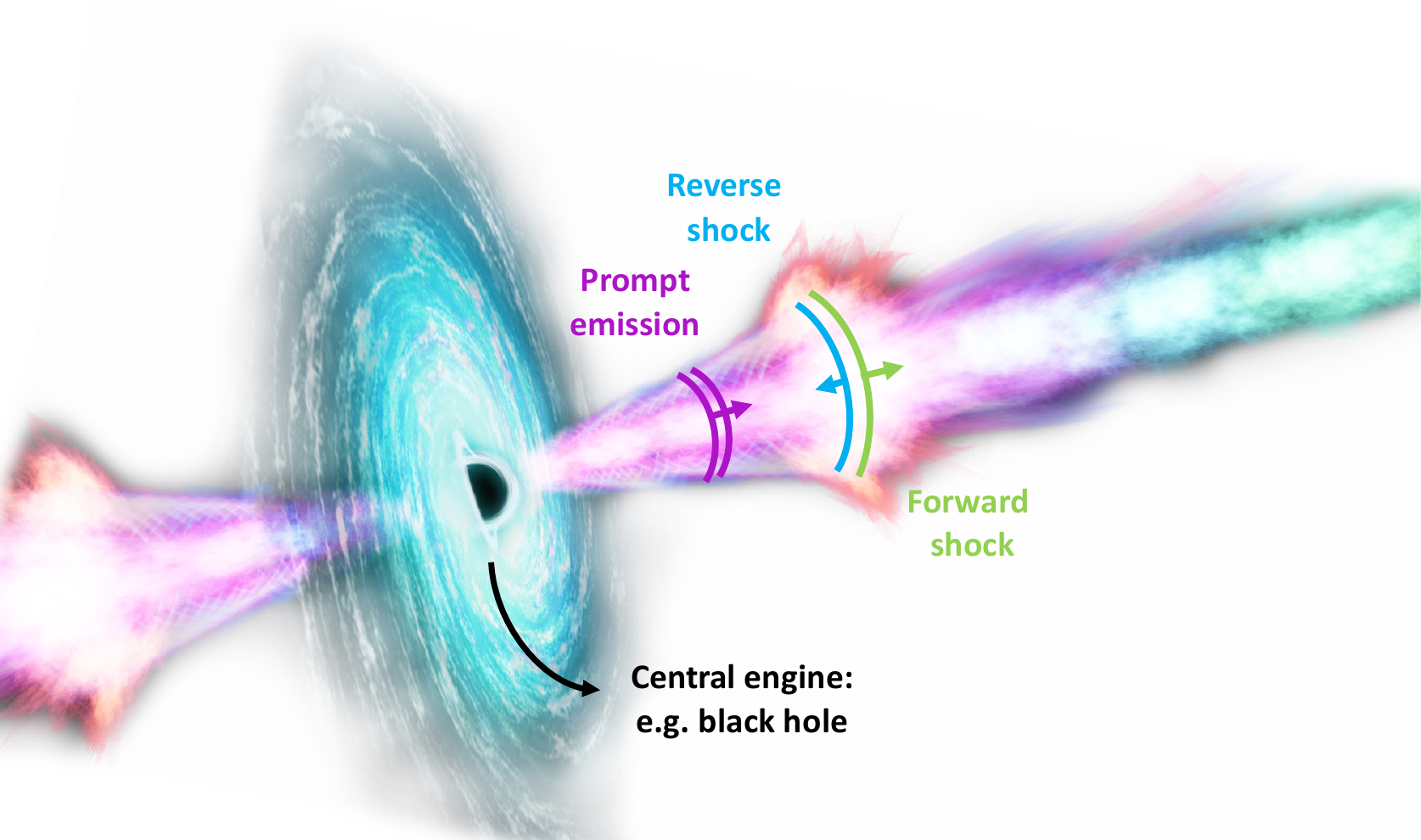}
\caption{Schematic representation of a GRB outflow. The steps for the formation of a GRB are: (1) an extragalactic massive star collapses or two compact objects merge, (2) there is accretion onto a newborn black hole, and (3) the ejecta are collimated and accelerated to relativistic speeds along the narrow beam of a jet. Note that the emission we detect (prompt gamma-ray emission and afterglow) comes from sites at different distances from the central engine. Additionally, the afterglow (reverse and forward shock) is formed at the contact discontinuity between the outflow and the circumburst medium.}
\label{fig:GRB_scheme}
\end{figure}

GRBs are produced when the central engine (a black hole or magnetar) drives out two bipolar relativistic jets that burrow through the stellar ejecta and emit the characteristic prompt gamma-ray emission via internal dissipation mechanisms\citep{1997ApJ...485..270S} ---i.e. internal shocks\citep{1994ApJ...430L..93R} or reconnection\citep{2011ApJ...726...90Z}. Later on, the relativistic ejecta are decelerated by the circumburst medium by a pair of external shocks\cite{1997ApJ...489L..37S, 1999PhR...314..575P}: a short-lived reverse shock \citep{1999ApJ...520..641S,2000ApJ...545..807K} and a forward shock (see Fig.~\ref{fig:GRB_scheme}). This lagging emission called the afterglow radiates via synchrotron emission and can be detected seconds to years after the burst at wavelengths across the electromagnetic spectrum\citep{1997Natur.387..783C,1997Natur.386..686V}. In this context, a variety of afterglow light curves are expected depending on the relative contributions of the reverse and forward shock.

\section{Magnetic fields in GRBs}

Magnetic fields play an important role in the physics of GRBs by affecting how the jet is launched (formation, acceleration, collimation and composition of the jet) and in shaping the afterglow emission (particle acceleration by the shock). Depending on the driving mechanism of the relativistic outflow ---if dominated by kinetic or magnetic energy--- GRBs jets can be launched as baryonic (also named hydrodynamical jets\cite{1990ApJ...365L..55S}) or magnetized jets (i.e., Poynting-flux jets \cite{1992Natur.357..472U,2009MNRAS.394.1182K}). The degree of magnetization of an outflow is usually characterized by the parameter $\sigma$, which is defined as the ratio of magnetic to kinetic energy flux\cite{2005ApJ...628..315Z}.

In a baryonic jet ($\sigma \ll 1$), neutrino annihilation at the polar region of a hyper-accreting system would launch a thermally-driven fireball that accelerates the material\cite{1993ApJ...405..273W,2005A&A...436..273A}. In this model, tangled magnetic fields are locally generated in shocks \citep{1999ApJ...526..697M}.

In a magnetized jet ($\sigma \gg 1$), large-scale magnetic fields extract rotational energy from the black hole, accretion disk or magnetar (in a magnetohydrodynamic extraction \cite{1977MNRAS.179..433B,1982MNRAS.199..883B,1992Natur.357..472U}), which accelerates the material. The magnetization parameter $\sigma$ influences the dynamics of the jet and it is thought to decrease with increasing distance from the central engine \cite{2014IJMPD..2330002Z}. In this scenario, the magnetic field configuration is likely toroidal far from the black hole \cite{2001AandA...369..694S}. The detection or non-detection of this large-scale magnetic field remnant is crucial to support or discard the magnetic jet model. Furthermore, a plausible scenario is that GRB outflows are launched with a combination of both hot-baryonic and cold-magnetic components, leading to a broad range of magnetization degrees $\sigma$ with consequences in the prompt and afterglow emission phase \citep{2005ApJ...628..315Z,2008A&A...478..747G,2009MNRAS.394.1182K}.

\section{The polarization of GRBs}

GRBs are extragalactic collimated sources that are not possible to spatially resolve. Polarization offers a tool to study the magnetic field properties and the jet geometry on length scales orders of magnitude smaller than what imaging can probe. Additionally, it can give an extra dimension of information on the jet physics and underlying emission processes and provide complementary information that can break degeneracies in the temporal and spectral evolution of the emission. Overall, the expected polarization signals in GRBs depend on the magnetic field topology (i.e. ordered or random depending on the jet model), the jet angular geometry\cite{2004MNRAS.354...86R,2020MNRAS.491.3343G}, and the emission mechanisms. While the emission mechanisms for the prompt emission are still under debate\cite{2009ApJ...698.1042T}, the afterglow emission is widely assumed to be synchrotron\cite{1998ApJ...497L..17S}. In the lab frame, synchrotron-emitted photons are linearly polarized \citep{1979rpa..book.....R}, with $P \approx 70\%$. However, the intrinsic polarization from synchrotron emission can be averaged out if the magnetic field topology is random.

During the afterglow stage, the forward shock emission is produced by shocked ambient medium, with tangled magnetic fields locally produced and amplified by e.g., a two-stream magnetic instability (Weibel instability\cite{1959PhRvL...2...83W,1999ApJ...526..697M}). Consequently, the forward shock emission is insensitive to the magnetic field structure of the original ejecta and it is expected to be unpolarized (see Section \ref{sec:CollisionShocks}). In contrast, the prompt and reverse shock emission is still sensitive to the properties of the central engine ejecta and measuring the polarization allows discriminating between competing jet models (see Section \ref{sec:JetModels}).

\subsection{The polarization of baryonic and magnetized jets} \label{sec:JetModels}

\begin{figure}[]
\centering
\includegraphics[width=110mm]{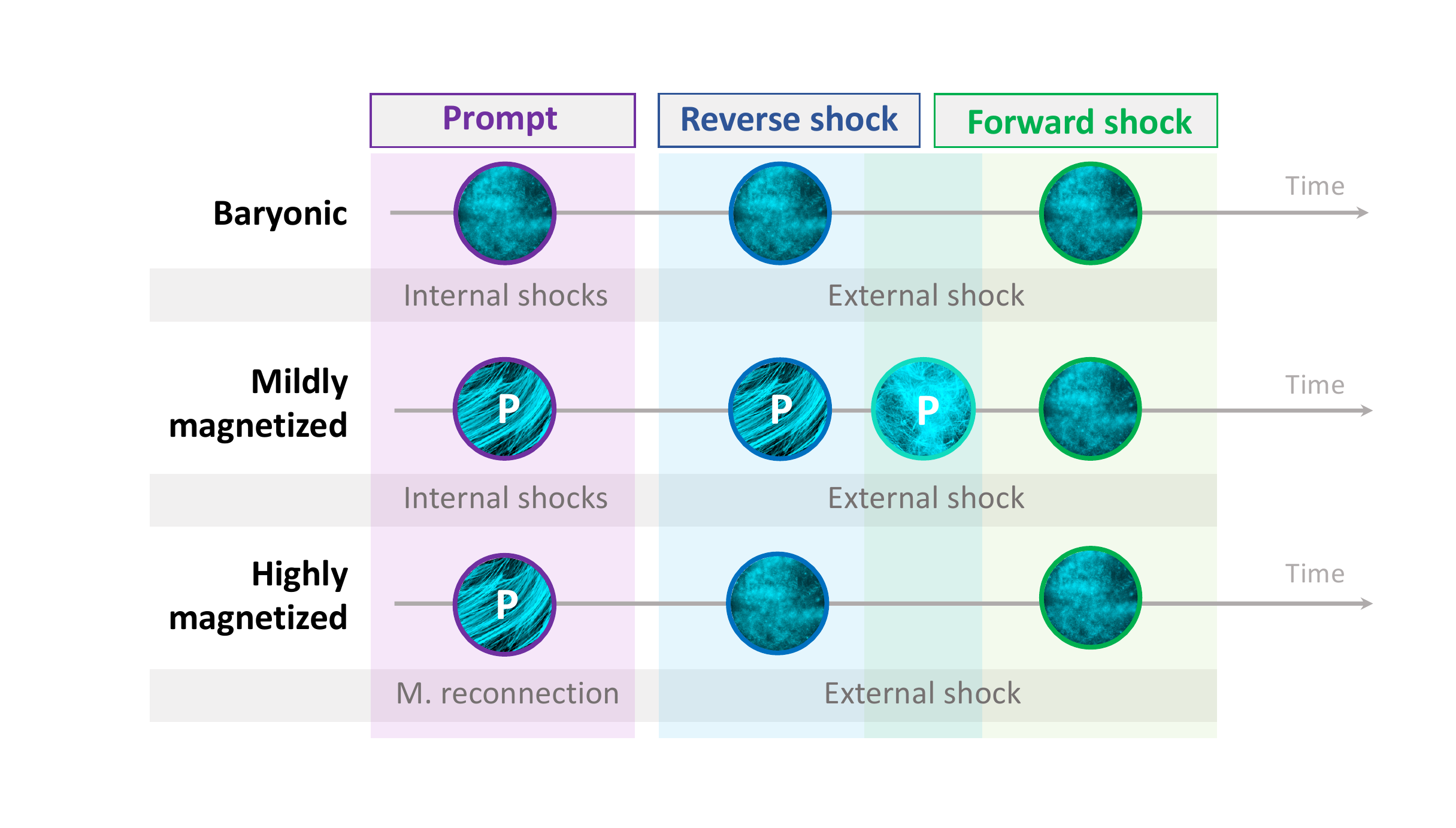}
\caption{Summary of the magnetic field structures expected in each jet model (baryonic and magnetized) for the different shocks (internal and external) and assuming synchrotron as the emitting mechanism. The white {\bf P} letter denotes those magnetic field structures that would give rise to significant polarization (large-scale ordered or locally tangled magnetic fields). Note that emission from different origins can be blended into the overall received; e.g. in a mildly magnetized jet, the polarization can be reduced from the maximum expected given a mix of reverse and unpolarized forward shock emission.}
\label{fig:baryonic_magnetized}
\end{figure}

Distinct polarization signatures are predicted for magnetized and baryonic jet models. We expect unpolarized emission if the jet is baryonic\citep{1999ApJ...526..697M} and polarized emission in a magnetized jet\citep{2003ApJ...594L..83G} from large-scale ordered magnetic fields advected from the central engine (see Fig.~\ref{fig:baryonic_magnetized}). Even if the magnetic fields become slightly distorted during the internal shocks, a mildly magnetized jet is still expected to produce large polarization levels, as only a small area of the jet is observed due to the relativistic beaming. However, note that the maximum synchrotron polarization is reduced to $P \approx 50\%$ due to relativistic aberration effects ---i.e., the rotation of the polarization vectors\cite{2003ApJ...597..998L}. In the case of a highly magnetized jet (see Fig.~\ref{fig:baryonic_magnetized}), reconnection mechanisms are thought to power the prompt emission\citep{2011ApJ...726...90Z,2016ApJ...821L..12D}, which would in principle distort the order in the magnetic fields.

\subsubsection{Prompt gamma-ray emission}

Polarization studies of the prompt gamma-ray emission offer great prospects for the study of jet models \cite{2020A&A...644A.124K}. However, current studies have yielded a wide range of polarization results\citep{2003Natur.423..415C, 2004MNRAS.350.1288R, 2004ApJ...613.1088W, 2009ApJ...695L.208G, 2013MNRAS.431.3550G,2012ApJ...758L...1Y, 2019ApJ...884..123C,2020A&A...644A.124K}, i.e. $P=0\%-100\%$, with the discrepancy usually related to data with low signal-to-noise, instrumental bias or selection effects in the data analysis\cite{2004MNRAS.350.1288R,2020A&A...644A.124K}. Furthermore, time-resolved analyses suggest that the polarization degree in time-integrated studies has been averaged out given that the polarization angle evolves through single pulses (see Refs.~\citenum{2019A&A...627A.105B,2019ApJ...882L..10S,2019NatAs...3..258Z}). More importantly, the interpretation of these results is complex given that the underlying physics of the prompt emission remain unknown\citep{2009ApJ...698.1042T}.

\begin{figure}[ht!]
\centering
\includegraphics[width=110mm]{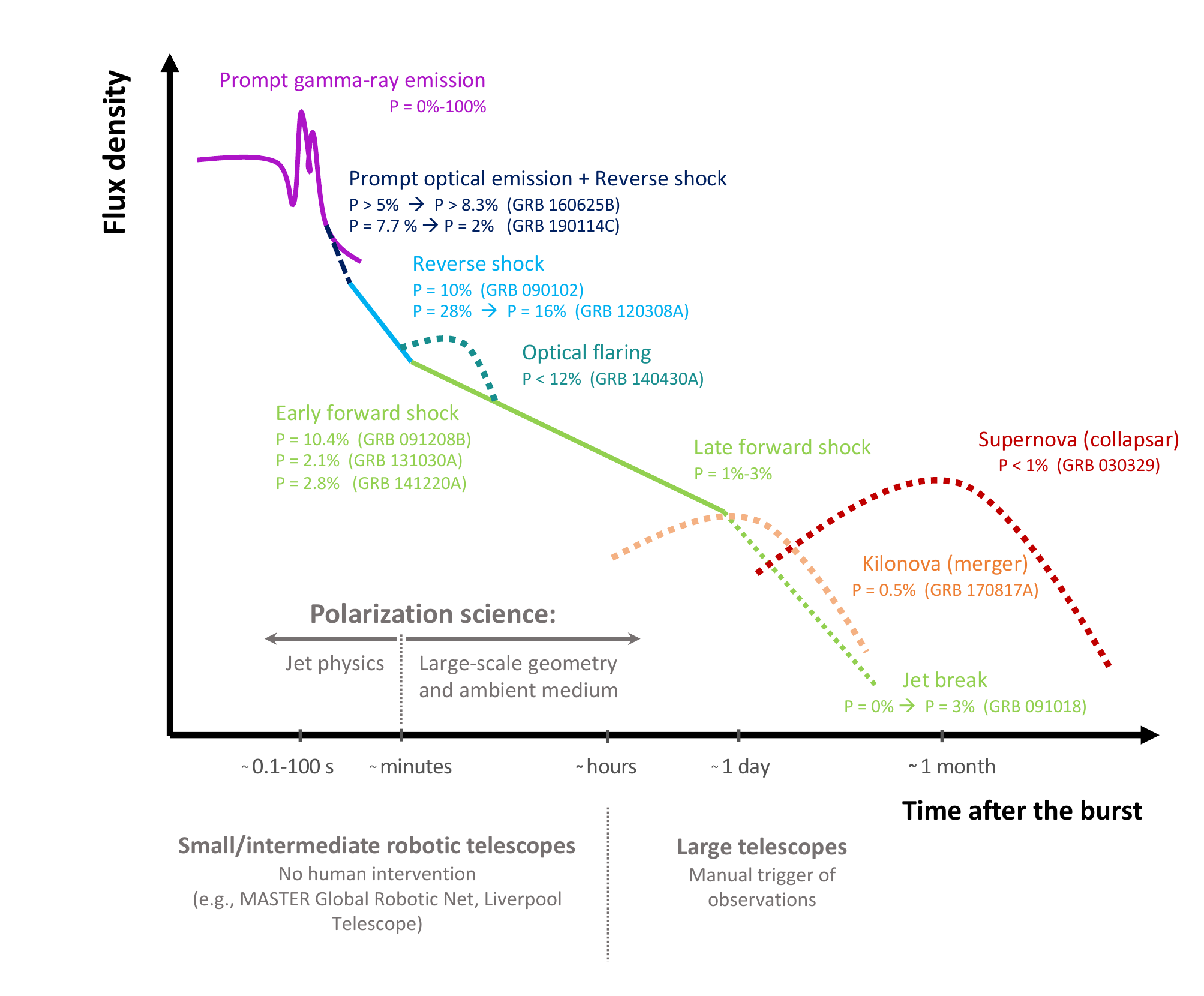}
\caption{Schematic representation of the different types of GRB emission and relevant polarization measurements. Note that, with the exception of the prompt gamma-ray emission, the represented "canonical" light curve refers to optical wavelengths.
{\it References:}
GRB 030329 \citep{2003ApJ...593L..19K}
GRB 090102 \citep{2009Natur.462..767S},
GRB 091018 \citep{2012MNRAS.426....2W},
GRB 091208B \citep{2012ApJ...752L...6U},
GRB 120308A \citep{2013Natur.504..119M},
GRB 131030A \citep{2014MNRAS.445L.114K},
GRB 140430A \citep{2015ApJ...813....1K},
GRB 141220A \cite{2021MNRAS.505.2662J}
GRB 160625B \citep{2017Natur.547..425T},
GRB 170817A \citep{2017NatAs...1..791C},
GRB 190114C\cite{2020ApJ...892...97J},
 late-time forward shocks\cite{2004ASPC..312..169C} and the AstroSat CZT/POLAR gamma-ray polarization catalogues \citep{2019ApJ...884..123C,2020A&A...644A.124K}.}
\label{fig:GRB_elements}
\end{figure}

\subsubsection{The early optical afterglow}

\begin{figure}[]
\centering
    \includegraphics[width=100mm]{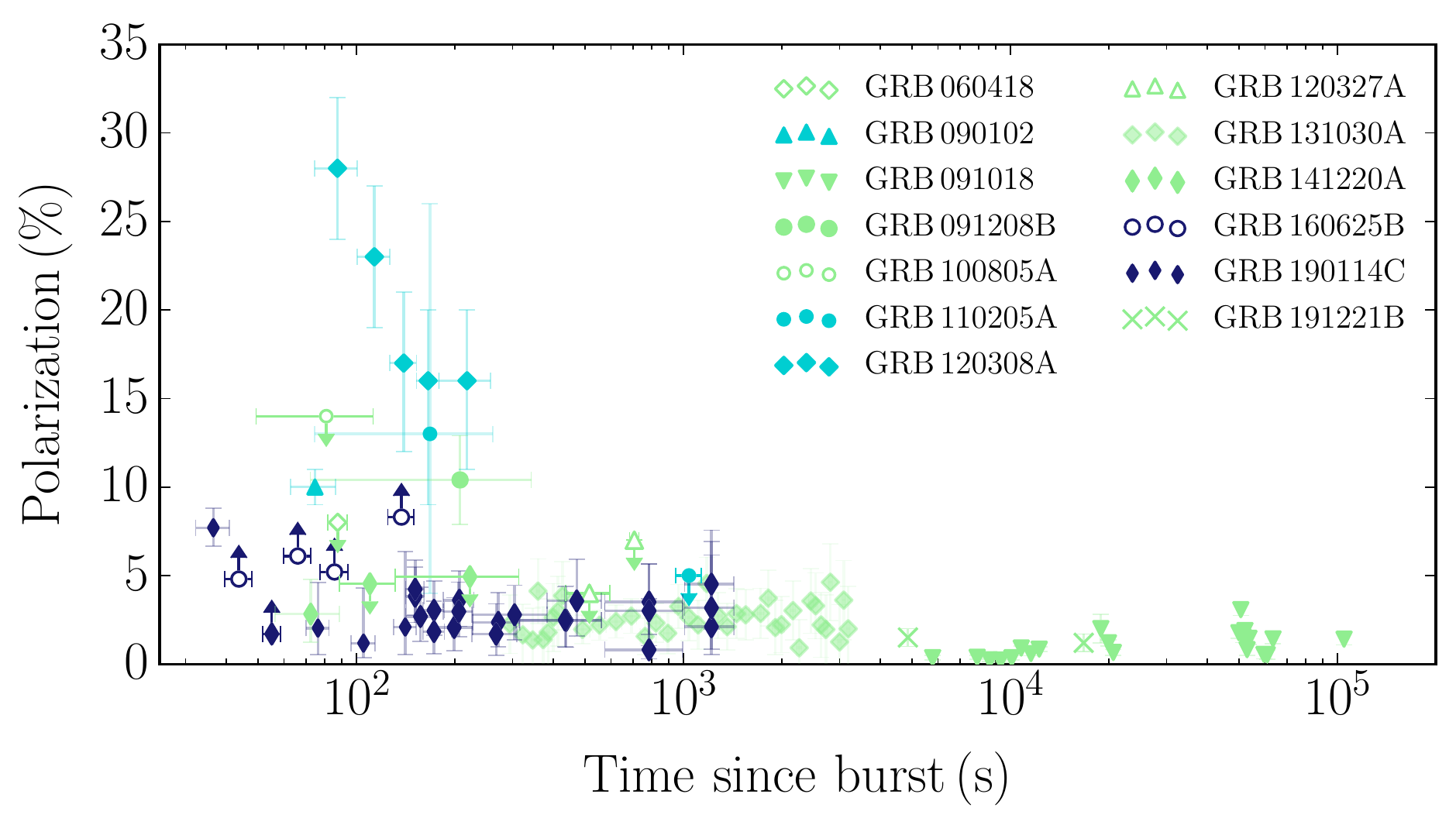}
\caption{Rest-frame polarization measurements of GRBs at optical bands ---adapted from Ref.~\citenum{2013Natur.504..119M}. Overall, the polarization dependency with time indicates that the regime in which we can infer the jet physics corresponds to observations before $\sim 5 \,$min postburst. In contrast, the polarization at later times gives information about the jet geometry and dust. {\it References:}
GRB 060418 \cite{2007Sci...315.1822M},
GRB 090102 \citep{2009Natur.462..767S}, 
GRB 091018 \citep{2012MNRAS.426....2W}, 
GRB 091208B \citep{2012ApJ...752L...6U},
GRB 110205A  \citep{2017ApJ...843..143S},
GRB 100805A \citep{2017ApJ...843..143S},
GRB 120308A \citep{2013Natur.504..119M}, 
GRB 120327A \citep{2017ApJ...843..143S},
GRB 131030A \citep{2014MNRAS.445L.114K},
GRB 141220A\cite{2021MNRAS.505.2662J},
GRB 160625B \citep{2017Natur.547..425T}, 
GRB 190114C\cite{2020ApJ...892...97J}, and
GRB 191221B\citep{2021MNRAS.506.4621B}. }
\label{fig:pol_all}
\end{figure}

At optical bands, the polarization measurements from the early afterglow are the domain of robotic telescopes with rapid follow-up capabilities\cite{2004AN....325..580L,2004AN....325..519S,2008SPIE.7014E..4LK,2019MNRAS.485.2355R} (see Fig.~\ref{fig:GRB_elements}). The first robust evidence of large-scale ordered magnetic fields in the fireball was the detection of  $P=10 \% \pm 1\%$ polarization during the steep decay of GRB 090102 bright reverse shock \citep{2009Natur.462..767S}. This result suggested that primordial magnetic fields do play a role in GRBs outflows and that they can persist at great distances from the central source ---up to the external shock radius. The GRB 120308A measurements further proved this idea with a time-resolved polarimetric light curve during a reverse-forward shock interplay \citep{2013Natur.504..119M} (see Fig \ref{fig:pol_all}); the polarization decayed steadily from $P=28\% \pm 4\%$ to $P=16^{+5} _{-4}\%$ with constant polarization angle due to the increase of unpolarized forward shock photons. Additionally, polarization degrees of $P=13^{+13} _{-9}\%$ and $P \approx 6\%$  were also detected in GRB 110205A and GRB 101112A\citep{2017ApJ...843..143S}.

Overall, current observations of polarized reverse shocks and the modelling of the early GRB afterglow suggest mildly magnetized jets at the deceleration radius\cite{2005ApJ...628..315Z,2014ApJ...785...84J}, with magnetization degrees $\sigma =0.1-1$. These intermediate values of magnetization allow the existence of bright reverse shocks and are also consistent with the measurements of polarized reverse shocks (see Fig.~\ref{fig:baryonic_magnetized}). However, we note that some observations suggest higher magnetization at different stages of the jet. The $P<8\%$ constraint during the fireball deceleration of GRB 060418\cite{2007Sci...315.1822M} implies that the reverse shock could have been suppressed due to high magnetization in the fireball at the external shock radius, with $\sigma \geq 1$. Additionally, the optical emission in between gamma-ray pulses of GRB 160625B \citep{2017Natur.547..425T} was interpreted as reverse shock and presented a significant increase from $P>5.2 \% \pm 0.6  \%$ to $P>8.3 \% \pm 0.5  \%$ simultaneous to a gamma-ray pulse ---implying that polarized prompt emission photons contributed at optical wavelengths. The polarization lower limits were quite low through observations, which led to speculations regarding distorting mechanisms during the prompt gamma-ray emission, such as reconnection\cite{2011ApJ...726...90Z,2016ApJ...821L..12D}.

\subsubsection{The highly magnetized jet of GRB 190114C}

Both MASTER II from the Master Global robotic net and RINGO3 instrument from the Liverpool Telescope observed GRB 190114C, the first GRB detected at very high energies (in the TeV domain\citep{2019Natur.575..455M}). Interestingly, the afterglow emission decay rate presented a steep to shallow transition, which suggested reverse shock contribution \cite{2020ApJ...892...97J}. Our observations report remarkably low polarization just after the end of the gamma-ray flash ($P=7.7\%$), a sharp drop of polarization one minute later ($P=2\%$) and constant levels during the following half an hour. In Ref.~\citenum{2020ApJ...892...97J}, we suggest that the $P\approx 2\%$ constant polarization is due to differential dust absorption in the line of sight ---mainly due to the highly obscured environment in which the GRB was formed--- which means that the intrinsic polarization of the jet is very small for reverse shock emission. Our low optical polarization measurements are also consistent with $P = 0.87 \% \pm 0.13 \%$ measured $2.2\,$h later at millimetre frequencies, during the reverse shock emission \citep{2019ApJ...878L..26L}.

The temporal and spectral modelling of GRB 190114C emission indicates a clear interplay between the reverse and forward shock and more magnetization in the reverse shock, which suggests the existence of a primordial large-scale magnetic field ejected from the rotating black hole. However, GRB 190114C polarimetric observations reveal that the magnetic field in the ejecta was mostly random oriented in space, which does not agree with what we expect from previous measurements of reverse shocks \citep{2009Natur.462..767S,2013Natur.504..119M} (see Fig.~\ref{fig:pol_all}).

In Ref.~\citenum{2020ApJ...892...97J}, we propose that the polarization was low because the large-scale magnetic field catastrophically collapsed during the first tens of seconds of the gamma-ray flash via magnetic reconnection mechanisms and that the $P=7.7\%$ measurement is a relic from this emission\citep{2011ApJ...726...90Z,2016ApJ...821L..12D}. As opposed to previous polarization measurements \citep{2009Natur.462..767S,2013Natur.504..119M}, these findings suggest that at least some energetic GRBs can be launched highly magnetized and that magnetic dissipation mechanisms can power a bright gamma-ray prompt. Additionally, it opens a debate about the nature of the bright early-time steep emission, which could be a reconnection tail instead of a reverse shock. That is because the reverse shock emission could have been suppressed if, at the deceleration radius, the outflow was still in the highly magnetized regime \cite{2008A&A...478..747G}. These results pin down timescales and distances for which large-scale magnetic fields survive in astrophysical jets and challenge the current models for the production of GRBs.

\subsection{The polarization from collisionless shocks} \label{sec:CollisionShocks}

Polarimetric studies of the late afterglow have measured low degrees of polarization at optical wavelengths during the decay of the long-lived forward shock\cite{2004ASPC..312..169C} and have provided information on the angular structure of jets\cite{2004MNRAS.354...86R} and the dust properties of GRB host galaxies\cite{2004AandA...420..899K} (see Fig.~\ref{fig:pol_all}). Additionally, these polarization levels agree with theoretical predictions and suggest that forward shocks are intrinsically unpolarized due to tangled magnetic fields locally generated in shocks \citep{1999ApJ...526..697M}.

\begin{figure}[]
\centering
	\includegraphics[width=125mm]{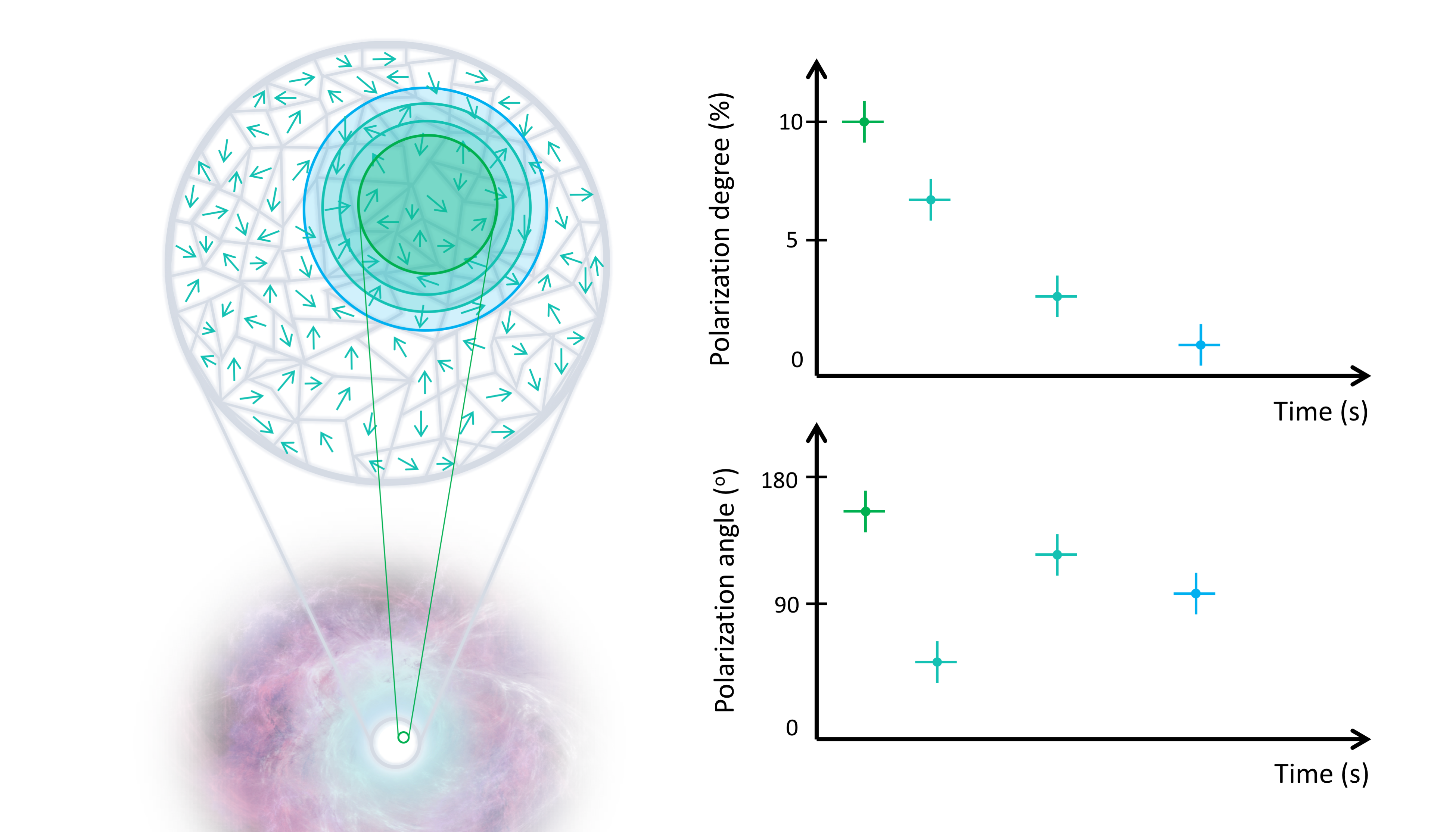}
    \caption{Schematic representation of the evolution of the polarization degree and angle in the patches model \citep{1999ApJ...511..852G}.}
    \label{fig:patches}
\end{figure}

However, Ref.~\citenum{2012ApJ...752L...6U} observations set a debate on the intrinsic polarization of forward shocks at early times given the detection of significant polarization in the GRB 091208B forward shock ($P=10.4\% \pm 2.5\%$), which favoured the patches model \citep{1999ApJ...511..852G}. Phenomenologically, the patches model implies that as the outflow decelerates and the visible emitting region increases, more patches with coherent magnetic fields (i.e., highly polarized; $P \approx 50 \%$) are visible. As the orientation of the magnetic field is random from one domain to another, the overall observed polarization decreases with time (see Fig.~\ref{fig:patches}). Consequently, GRB 091208B measurement still agrees with the $P \approx 1\% - 3 \%$ polarization found in late-time forward shocks. Physically, the patches model proposes that the amplification mechanism of the magnetic field at the shock front ---needed to increase the magnetic energy density from interstellar to the observed levels--- leads to the formation of large magnetic domains (i.e. large coherent length scales for the magnetic field). 

The GRB 091208B measurement gives rise to new possible mechanisms for magnetic field amplification, as well as adding a problematic degeneracy in early-time polarimetric studies. Note that the GRB 091208B polarization level is comparable to that measured in the reverse shock of GRB 090102 \citep{2009Natur.462..767S}. If forward shocks are polarized at early times, it becomes difficult to disentangle the possible polarized components that probe the jet physics from the contribution of the shocked ambient material. Consequently, a good understanding of the typical polarization levels of early-time forward shocks is crucial.

\subsubsection{The polarization of the early-time classical forward shock of GRB 141220A}

In Ref.~\citenum{2021MNRAS.505.2662J}, we report the early-time observations of GRB 141220A forward shock. Interestingly, this afterglow presented temporal and spectral properties typical of late-time forward shocks but starting as soon as 1.4 min postburst. Furthermore, our polarimetric observations at 2.2-3.4 min postburst are earlier than GRB 091208B measurement (2.5-11.8 min), which allow us to probe smaller emitting regions. If the length scale of the magnetic field was the same as in GRB 091208B, we would expect $P=20\%$ polarization for GRB 141220A at the time of observations (as predicted by the patches model; see Fig \ref{fig:patches}). Instead, we measure $P = 2.8 _{-  1.6} ^{+  2.0} \, \%$ in GRB 141220A, which we find is compatible with that induced by dust in the GRB environment and agrees with theoretical predictions of unpolarized forward shocks\citep{1999ApJ...526..697M, 2011ApJ...734...77I}, i.e. $P_{\rm  th} = 0\%- 2\%$ (but see Refs.~\citenum{2013ApJ...776...17M,2017ApJ...838...78M}).

The GRB 141220A and current early-time forward shock polarization measurements suggest that GRB 091208B is an outlier (see Fig.~\ref{fig:pol_all}). We speculate that GRB 091208B scenario could be similar to the reverse-forward shock interplay in GRB 120308A \citep{2013Natur.504..119M}. As the polarization measurement was integrated over a large time window, we suggest the presence of a more stable underlying polarized component and ordered large-scale magnetic fields in the fireball.

\section{Outlook}

After 15 years of progress in early-time polarization studies of GRBs afterglows, we can confidently confirm the presence of large-scale ordered magnetic fields in a subset of bright afterglows, suggesting that the magnetic field ejected from the central source does play a role in the jet launching. In general, we have not measured the maximum polarization allowed by theory in reverse shocks ($P=50\%$) and polarization observations indicate that highly polarized emission at $P=28\%$ level is not that common (GRB 120308A\cite{2013Natur.504..119M}; see Fig.~\ref{fig:pol_all}). Therefore, mechanisms that can change the magnetic field topology might play an important role at the early stages of GRB jets.

Observations of reverse shocks remain rare \citep{2014ApJ...785...84J}, and a way to overcome the lack of reverse shocks in the optical is to look at millimetre and radio wavelengths. The predictions are that, if the forward shock peaks at optical bands, the reverse shock will peak at $1-100 \,$GHz bands and $\sim 0.01-0.1 \,$days postburst with fluxes of $0.01-0.1 \,$mJy \citep{2015ApJ...806..179K}. However, if the magnetization is high at the external shock radius, the reverse shock would be suppressed \citep{2008A&A...478..747G}. The lack of reverse shocks at optical and radio wavelengths would indirectly imply that jets are still highly magnetized at large distances from the central source. Consequently, the rate and the polarization of reverse shocks at optical \citep{2009Natur.462..767S,2013Natur.504..119M,2017ApJ...843..143S,2017Natur.547..425T,2020ApJ...892...97J} and lower frequencies \citep{2019ApJ...878L..26L} can greatly constrain the magnetization of GRB jets and further test the mildly and highly magnetized jet regimes.

Great advancement will be made on constraining GRB jet models with a faster target acquisition and the increase of the polarimeters' sensitivity \cite{2019AN....340...40G}. This will allow gathering a large polarimetric dataset of measurements at early times, sampling most of the GRB population at lower luminosities. Furthermore, other TeV GRBs have been discovered\citep{2019Natur.575..464A,2020GCN.29075....1B,2021ApJ...908...90A,2021Sci...372.1081H} ---something we did not expect before the Cherenkov Telescope Array (CTA). With the start of the CTA era, we will be able to track the earliest stages of the magnetic field and its evolution in the most energetic systems. Consequently, we will need new polarization technology to optimize CTA follow-up as well as other multimessenger triggers from gravitational waves and possibly neutrinos.

\bibliographystyle{ws-procs961x669}
\bibliography{biblio}

\end{document}